\documentclass[sigconf]{acmart}

\usepackage{graphicx}

\usepackage{subcaption}
\usepackage{cleveref}
\usepackage{booktabs}
\usepackage{multirow}
\usepackage{hyperref}
\usepackage{lipsum}
\usepackage{adjustbox}
\usepackage{tcolorbox}
\usepackage[table]{xcolor}
\usepackage{pgfmath}

\newcommand{\blua}[1]{\cellcolor{blue!5}#1}
\newcommand{\blub}[1]{\cellcolor{blue!15}#1}
\newcommand{\bluc}[1]{\cellcolor{blue!30}#1}
\newcommand{\blud}[1]{\cellcolor{blue!50}#1}
\newcommand{\blue}[1]{\cellcolor{blue!75}#1}

\title{Is~Three~the~Magic~Number? An~Empirical~Evaluation~of~LLM-Based~Repair~Loops}

\author{Tobias Kiecker}
\affiliation{
  \institution{Humboldt-Universität zu Berlin}
  \city{Berlin}
  \country{Germany}
}
\author{Eik Reichmann}
\affiliation{
  \institution{Humboldt-Universität zu Berlin}
  \city{Berlin}
  \country{Germany}
}
\author{Hosung Kang}
\affiliation{
  \institution{Korea University}
  \city{Seoul}
  \country{Korea}
}
\author{Gabin An}
\affiliation{
  \institution{Korea University}
  \city{Seoul}
  \country{Korea}
}
\author{Lars Grunske}
\affiliation{
  \institution{Humboldt-Universität zu Berlin}
  \city{Berlin}
  \country{Germany}
}
\setcopyright{none}
\renewcommand\footnotetextcopyrightpermission[1]{}

\acmConference{Pre-Print}{arXiv}{Germany}
\acmBooktitle{PrePrint}
\acmYear{}
\copyrightyear{}

\begin{abstract}
Iterative repair loops have become a core design pattern in LLM-based software engineering systems. 
These workflows repeatedly generate, validate, and repair artifacts using feedback such as compiler errors or test failures. 
Despite their widespread use, the impact of repair-loop iteration limits remains poorly understood, as most prior work adopts fixed, often arbitrary, repair budgets.

We study repair-loop effectiveness across multiple software engineering tasks, including code generation, test generation, and code translation. 
Across several representative workflows, datasets, and contemporary low-cost LLMs, we observe a consistent pattern of diminishing returns: the first three to four repair iterations account for most achievable gains, while later iterations contribute only marginal improvements. 
We further find that repair behavior is influenced more strongly by workflow orchestration and feedback design than by the underlying model itself. 
These results suggest that repair budgets should be treated as an explicit experimental variable, as they directly affect evaluation outcomes, computational cost, runtime, and reproducibility in LLM-based software engineering research.

\end{abstract}

\begin{document}

\maketitle

\section{Introduction}\label{sec:intro}
Large Language Models (LLMs) are far from perfect.
Nevertheless, they are increasingly used in software engineering (SE) research and practice~\cite{fan2023large}.
A common pattern when using LLMs for programming tasks is iterative refinement: users generate a solution, validate the output, identify remaining issues such as compiler errors, failing test cases, or missing edge cases, and feed these back to the model for repair. This process is repeated until the generated artifact either satisfies the requirements or the user abandons the attempt and reformulates the prompt.
This iterative repair behavior has also been adopted by many LLM-based SE tools~\cite{tool:selfrepair, tool:alphacodium, tool:aster, tool:cascade, tool:CoT, tool:intervenor, tool:ldb, tool:PairCoder, tool:repocoder, tool:self-refine}. 

Instead of relying on manual feedback, such systems automatically validate generated artifacts using mechanisms such as compilation, static analysis, or test execution. When validation fails, the observed errors are provided to the model in order to generate a repaired version automatically. To prevent infinite repair cycles or oscillating behaviors in which one repair introduces new faults, these systems typically enforce an arbitrary fixed iteration limit~\cite{tool:oci, tool:testspark}. 
Furthermore, with the emergence of autonomous LLM agents~\cite{zhang2024autocoderover,bouzenia2025repairagent,yang2024swe} repair loops are becoming even more relevant. 
Such agents internally perform iterative cycles of tool calls, generation, validation, testing, and repair in order to accomplish complex SE tasks. 
As agent-based workflows become increasingly common, the choice of repair-loop limits directly affects the practicality and efficiency of these systems.

Choosing the right limit for repair loop iterations is important and impactful for several reasons. 
First, each additional repair attempt increases API usage costs and energy consumption, contributing both to higher economic as well as environmental impact~\cite{EnergyCosts}. 
Second, repeated inference calls increase runtime which can lead to slower development workflows.  
Third, iteration limits directly influence reproducibility and comparability across studies. Different repair budgets may lead to substantially different performance outcomes, making it difficult to fairly compare approaches when iteration limits are selected arbitrarily or are insufficiently documented~\cite{GaoHGXJ25}. 

\begin{figure}[t]
    \centering
    \includegraphics[width=0.85\linewidth]{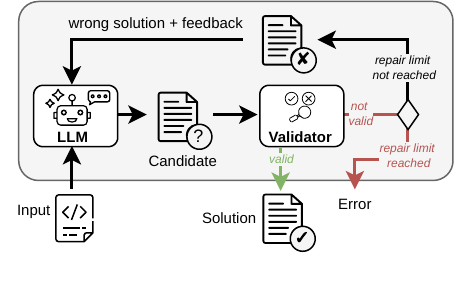}
    \vspace{-2\baselineskip}
    \caption{Generalized SE workflow with LLM-repair loop}
    \label{fig:repairloop}
    \vspace{-1\baselineskip}
\end{figure}

Despite their importance, iteration limits are rarely studied systematically. Existing work often selects a small fixed number of repair attempts without providing empirical justification for this choice~\cite{tool:oci,tool:CoT,tool:self-refine}.
To the best of our knowledge, no prior work systematically studies the impact of repair-loop iteration limits across SE tasks and LLMs.
In this work, we investigate the effectiveness of repair loops across multiple SE tasks, including code generation, test generation, and code translation. Using several low-cost LLMs and representative LLM-based SE tools, we analyze whether the commonly used limit of three or five repair attempts is sufficient or whether meaningful improvements continue to occur beyond this threshold. 
Our findings aim to contribute to a better understanding of how repair budgets should be selected and reported in future LLM-based SE research and systems.

Our central question is: 

\begin{center}
\tcbox{
    \textit{Are 3 repair attempts enough for LLM-based SE tools?}
}
\end{center}

\section{Repair Loops in LLM-Based SE-Tools}

We define a \emph{repair-loop}-based system as any LLM-based software engineering workflow, tool, or approach that iteratively generates, validates, and repairs artifacts until either a correct solution is produced or a predefined iteration limit is reached. \Cref{fig:repairloop} illustrates the general structure of such an approach.

These systems typically receive input in the form of task-specific context and an initial generation prompt. For example, a system may receive a textual specification and is prompted to generate source code, or receive source code and is prompted to generate corresponding test cases. 
Based on this input, the LLM generates an initial candidate artifact. The generated output is then validated using task-dependent mechanisms. In code generation scenarios, this may involve compilation, static analysis, or test execution. In test-generation tasks, generated tests may be checked for syntactic validity, coverage, or execution behavior.

If the produced artifact satisfies the validation criteria, the workflow terminates successfully and returns the generated output. Otherwise, the system extracts feedback from the failed validation step and provides this information back to the LLM to generate a repaired version of the artifact. 

The repaired artifact is then validated again, resulting in an iterative generate $\rightarrow$ validate $\rightarrow$ repair cycle. 
This process continues until either a valid artifact is produced or a predefined iteration limit is reached and an error is returned. 
In practice these iteration limits are commonly implemented implicitly through bounded loops or explicitly through retry counters within the orchestration logic of the system. Despite being a central configuration parameter, these limits are often selected heuristically and rarely justified empirically~\cite{tool:aster, tool:intervenor, tool:CoT}.

In the context of this work, \textit{repair} generally refers to feedback-driven regeneration conditioned on previous validation failures and we define a \emph{repair step} as one iteration of the repair loop following a failed validation attempt.

\section{Empirical Study}

To answer our posed question we conduct an empirical study across a set of tools with different tasks, datasets, and programming languages. The tasks span code generation, test-case generation and code translation.
To ensure that any observed results are not due to a particular model's behavior we select three models: Google's Gemma-4\footnote{\href{https://huggingface.co/google/gemma-4-31B-it}{huggingface.co/google/gemma-4-31B-it}}, Qwen3.5\footnote{\href{https://huggingface.co/Qwen/Qwen3.5-27B}{huggingface.co/Qwen/Qwen3.5-27B}} and OpenAI's \texttt{gpt-4o-mini}\footnote{\href{https://platform.openai.com/docs/models/gpt-4o-mini}{platform.openai.com/docs/models/gpt-4o-mini}}.

We study four existing repair-based tools and a baseline that simulates human-style iterative chat interactions solving coding problems. The tools include INTERVENOR~\cite{tool:intervenor}, Large Language Model Debugger (LDB)~\cite{tool:ldb}, from which we use the code translation component, CASCADE~\cite{tool:cascade} and OpenCodeInterpreter (OCI)~\cite{tool:oci} in two versions. First, the original version where the initial prompt contains all test-cases that should pass and a second version without test-cases in the initial prompt. \Cref{tab:overview} lists the tools and their datasets. Each datasets was also used in the tools original papers. For the baseline (Basic LLM Call) we use Defects4J~\cite{dataset:defects4j}.

\begin{table}[t]
   \centering
   \caption{Tools, datasets, and tasks used in our empirical study.}
   \label{tab:overview}
\begin{tabular}{@{}lll@{}}
\toprule
Tool & Dataset & Task \\ \midrule
Basic LLM Call & Defects4J \cite{dataset:defects4j} & Test Generation \\
OCI \cite{tool:oci} & MBPP \cite{dataset:mbpp} & Code Generation \\
INTERVENOR \cite{tool:intervenor} & HumanEval \cite{dataset:humaneval} & Code Generation \\
CASCADE \cite{tool:cascade} & CASCADE \cite{tool:cascade} & Test Generation \\
LDB \cite{tool:ldb} & TransCoder \cite{dataset:transcoder} & Code Translation \\ \bottomrule
\end{tabular}
\end{table}

To enable a comparison between the tools, we made two types of modifications. First, all tools were adapted to support the backbone LLMs that were hosted on a local server (Gemma-4 and Qwen3.5). For INTERVENOR 
this entailed a change from the OpenAI \texttt{completions} to the \texttt{chat.completions} API to enable the usage of the contemporary models since the \texttt{completions} API is deprecated. OCI was restructured to use a multi-turn conversation format rather than rebuilding the prompt from scratch each iteration, as in the original. These changes slightly alter the semantics of OCI and INTERVENOR, however, the change to the API call would be needed anyways to use contemporary models and were not specific to our study. 

Second modification was to LDB, where the original implementation only evaluated correctness when the final repair loop is finished. We added per-iteration evaluation against the full held-out test suite at every step.
For CASCADE, whose primary purpose is detecting inconsistent documentation-method pairs, we extracted only its test-case generation stage and discarded the later inconsistency analysis stages, since the initial test-generation component is the only part that implements a complete iterative repair loop suitable for our study.
Crucially, the repair loop control logic and correctness criteria remained unchanged in all tools. The only changes were with respect to measurement and model compatibility. Furthermore, we fixed the temperature for all experiments at 0.2.

\begin{figure*}[t]
    \centering
    \includegraphics[width=0.99\linewidth]{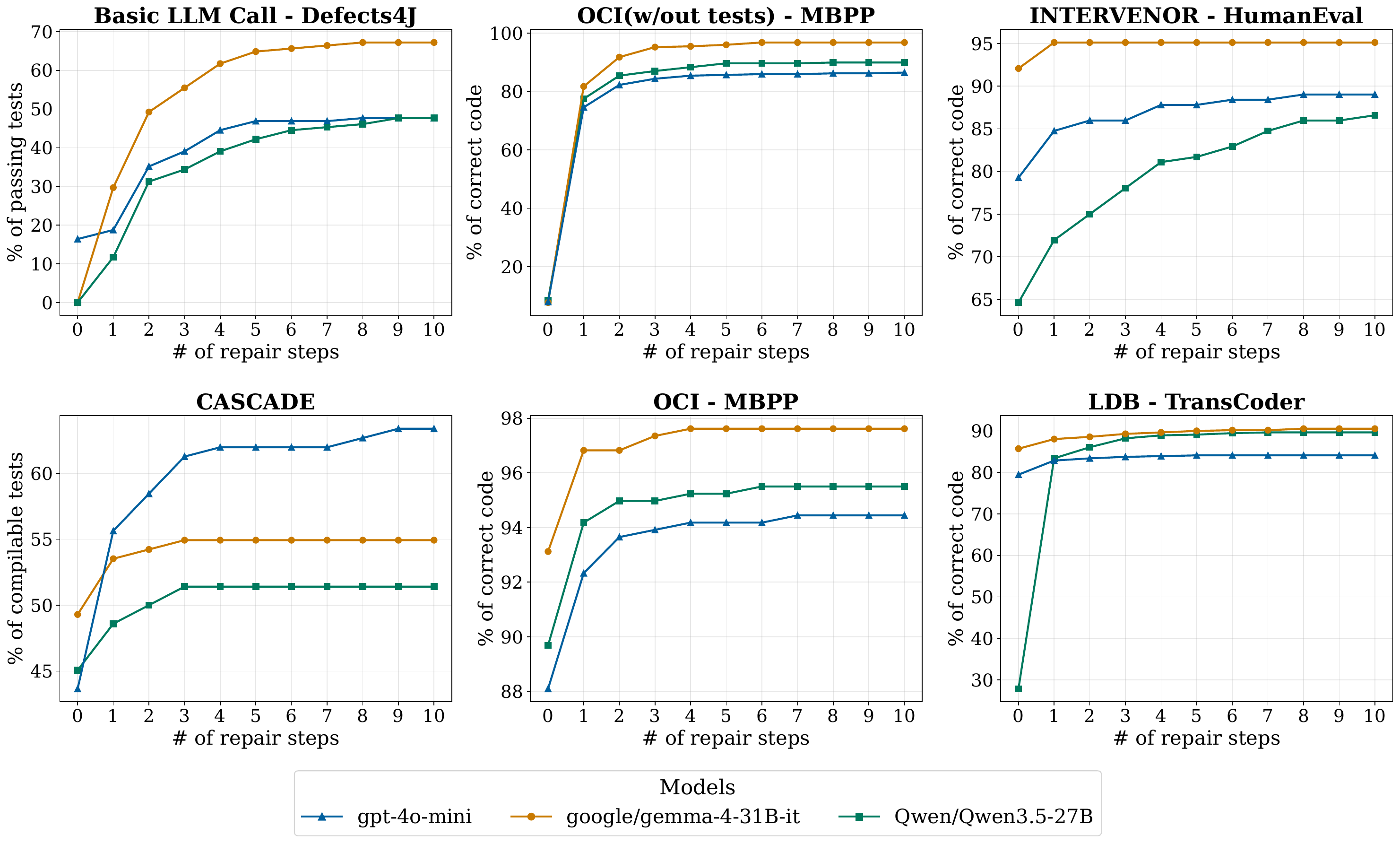}
    \caption{Completion rate as a function of repair steps across six tool–dataset combinations and three LLM backbones.}
    \label{fig:results}
\end{figure*}

For each tool with every backbone LLM we compute the percentage of completed tasks on the respective dataset after the initial generation, step 0, and at each subsequent repair step up to a maximum of ten repairs. Those results are shown in \Cref{fig:results}, where each panel corresponds to one tool and each curve to a model.

Across all tools the curves share a consistent concave shape. From the initial generation to the first few repair steps the completion rate rises steeply and then flattens, with each additional step resulting in smaller improvements than the last. \Cref{tab:imporvments} shows this in concrete numbers, where we calculated the average relative improvements from each repair step to the next. Notably, the largest relative improvement occurs at step 1 and that by step 3 the marginal gains have dropped to single digit percentages or below. This pattern holds across all three backbone models and all tasks including outliers such as Qwen on LDB starting with only 30\% completion rate and then improving drastically.
This suggests that diminishing returns are a recurring characteristic of the evaluated repair-loop workflows.

\begin{table}[b!]
   \centering
   \caption{Mean relative improvement per repair step.}
   \label{tab:imporvments}
\resizebox{\columnwidth}{!}{
\begin{tabular}{lrrrrrrrrrr}
\toprule
& \multicolumn{10}{c}{Percentage increase to previous step}           \\
                           & 1     & 2    & 3    & 4    & 5   & 6   & 7   & 8   & 9   & 10  \\ \midrule
Basic LLM Call& \blue{266.7} & \blud{92.2} & \bluc{11.5} & \bluc{12.7} & \blub{5.9} & \blub{2.0} & \blua{1.0} & \blub{1.5} & \blua{1.0} & 0.0 \\
INTERVENOR    & \bluc{6.7}   & \blub{1.7}  & \blub{1.2}  & \blub{1.9}  & \blua{0.2} & \blua{0.7} & \blua{0.7} & \blua{0.7} & 0.0 & \blua{0.2} \\
CASCADE        & \bluc{14.3}  & \blub{3.1}  & \blub{3.0}  & \blua{0.4}  & 0.0 & 0.0 & 0.0 & \blua{0.4} & \blua{0.4} & 0.0 \\
OCI         & \blub{4.6}   & \blua{0.7}  & \blua{0.3}  & \blua{0.3}  & 0.0 & \blua{0.1} & \blua{0.1} & 0.0 & 0.0 & 0.0 \\
OCI (w/out tests)  & \blue{871.4} & \bluc{11.0} & \blub{2.8}  & \blub{1.0}  & \blua{0.8} & \blua{0.4} & 0.0 & \blua{0.2} & 0.0 & \blua{0.1} \\
LDB        & \blud{31.7}  & \blub{1.5}  & \blub{1.2}  & \blua{0.5}  & \blua{0.3} & \blua{0.2} & \blua{0.1} & \blua{0.1} & 0.0 & 0.0 \\ \bottomrule
\end{tabular}}
\end{table}

When examining where tools differ, the main distinction lies in how quickly the return on repair diminishes. OCI shows only marginal improvements after the first repair step, while the Basic LLM Call and CASCADE are still rising at step 3. Furthermore, within each panel, models may start at different absolute levels but follow the same curve shape. The gap between models is largest at step 0 and narrows with each repair step, visible most clearly in LDB where Qwen starts near 30\% and Gemma near 85\% yet both curves flatten at roughly the same point. Furthermore, we can observe that each tool, independent of the model backbone, provides a different performance ceiling and a different rate at which repairs approach it.

\section{Discussion}
The first three to four repair steps contribute the vast majority of the total achievable gain. Following iterations can still repair additional cases, however, improvements drop to near zero shortly after steps 5 to 7 across all evaluated tasks and tools. Hence we argue that the costs that come with these improvements require a careful trade-off analysis with respect to parameters such as API costs and energy usage. This consideration is increasingly relevant for agent systems, where repair loops are often deeply integrated into the execution workflow and may trigger multiple expensive inference calls per task or even result in infinite loops.

Interestingly, our results indicate that factors such as feedback design and content, orchestration logic, and the validation strategy substantially influence how effective additional repair attempts are at improving generated artifacts. This comes from the observation that completion curves for the same tool have similar shapes across all models, even when the absolute performance levels differ. 

The comparison between the two OCI variants particularly highlights this effect. While both variants share nearly identical repair behavior in later iterations, they differ substantially in their initial success rates. The primary difference between the two approaches is that one variant already includes tests during the initial generation step, whereas the other only incorporates the output of the test suite as repair feedback during subsequent iterations. Despite this relatively small architectural difference, the resulting repair trajectories differ noticeably during the early repair stages before converging toward similar trends. 

At the same time, model-specific performance characteristics remain visible across tools. For almost all evaluated workflows, the Gemma model achieves the strongest overall performance, while the older GPT variant generally performs worst. However, CASCADE represents a notable exception to this trend. One possible explanation is that CASCADE was originally evaluated and probably optimized around the specific output of GPT-based models. 
This observation further reinforces that the interaction between tool repair logic and model behavior needs to be examined carefully.

Furthermore, our replication effort revealed that many existing LLM-based SE tools are not easily transferable across models. In practice, substantial modifications to prompts, parsing logic, validation mechanisms, or orchestration strategies are often required when replacing the originally intended backbone model. This raises broader concerns regarding comparability and reproducibility in LLM-based SE research. While a detailed investigation of reproducibility challenges is outside the scope of this work, our findings indicate that LLM tools and specifically repair-loop behavior cannot always be studied independently of the surrounding tool architecture and model assumptions and therefore cannot always be compared directly.

\section{Threats to Validity}

\paragraph{Internal Validity}
Our study does not use the original models employed by the evaluated tools and prior work. In addition, due to computational and financial constraints, we do not perform repeated executions for every experimental configuration. Since LLMs exhibit non-deterministic behavior, repeated runs may produce variance in absolute performance values.
However, the goal of this work is not to reproduce exact performance numbers, but to analyze repair-loop trends across different iteration limits. We therefore use the same set of models consistently across all evaluated tools and tasks.

Furthermore, the evaluation generally relies on task-specific automated validation procedures such as compilation and test execution.
While these mechanisms provide scalable correctness signals, they may not fully capture semantic correctness or overall software quality, however, they ensure consistency with existing evaluation methodologies.

\paragraph{External Validity}
Our experiments focus primarily on cost-efficient LLMs, and the observed results may therefore not generalize directly to larger, newer models. More capable models may require fewer repair attempts due to stronger reasoning and generation capabilities.

Furthermore, the selected tasks 
may not represent the full spectrum of repair-loop-based SE workflows. Nevertheless, we evaluate three substantially different SE tasks across five datasets, providing diversity in both generation objectives and validation mechanisms. Code generation in particular represents one of the most widely studied and practically relevant applications of LLMs in SE, increasing the practical relevance of the evaluated tools and benchmarks. The consistency of observed trends across multiple tasks and datasets partially mitigates this threat even further.

\section{Related Work}

Prior work on LLM-based repair can be categorized into three types. The first category is feedback driven generation that only uses an LLM and no information from the external executions~\cite{tool:self-refine,tool:repocoder}.

The next category characterizes tools that utilize some form of execution based feedback~\cite{tool:CoT}, including runtime and compilation errors~\cite{tool:oci,tool:aster,tool:testspark}, natural language~\cite{tool:alphacodium} and execution graphs~\cite{tool:ldb}.

Another category distributes the repair process across multiple agents or structured components. INTERVENOR~\cite{tool:intervenor} separates a Code Teacher, which constructs a Chain-of-Repair from compiler feedback, from a Code Learner, which executes the revisions, within three turns.
PairCoder~\cite{tool:PairCoder} combines planning and implementation agents with execution-guided repair within ten iterations.

Finally, our work is also related to empirical studies of repair and self-correction. Critical surveys note that self-correction effectiveness depends strongly on the feedback source and evaluation design, and that unclear protocols can overestimate gains~\cite{self-correction-critical}. Closest to our motivation, Olausson et al.~\cite{tool:selfrepair} compare repair with drawing additional independent samples under comparable compute budgets, while Tang et al.~\cite{tang2024code} frame repair as an exploration--exploitation tradeoff. In contrast, we ask a complementary question: Given that many LLM-based SE tools already implement repair loops, how much does each additional iteration help? We therefore measure per-iteration completion rates across existing tools, tasks, datasets, and model backbones, shifting the focus from whether to use repair to how iteration limits should be selected, justified, and reported.

\section{Conclusion}
Repair loops have become a fundamental component of modern LLM-based software engineering systems, ranging from code-generation tools to autonomous agent workflows. Despite their widespread use, the choice of repair-loop iteration limits is often treated as an arbitrary implementation detail and rarely justified empirically.

In this work, we conducted an empirical study across multiple repair-loop-based software engineering tools, tasks, datasets, and backbone models to analyze the effectiveness of additional repair iterations. Across all evaluated settings, we observe a consistent pattern of diminishing returns: the majority of achievable gains are obtained within the first three to four repair steps, while later iterations can still contribute marginal improvements.

Our results suggest that repair-loop behavior depends more strongly on orchestration logic, validation strategy, and feedback design than on the selected backbone model itself. At the same time, the observed interaction effects between tools and models indicate broader reproducibility and comparability challenges in LLM-based software engineering research.

Overall, our findings suggest that repair budgets should be explicitly reported and justified in future work, as they directly influence evaluation outcomes, computational cost, runtime, and environmental impact. As repair loops become increasingly integrated into autonomous agent systems, understanding and standardizing repair-loop behavior will become even more important for the development of reliable and reproducible LLM-based software engineering workflows. For future work we plan to extend on our results by adding more tools, models and conducting a larger scale reproducibility study for LLM-based SE tools.

\clearpage
\bibliographystyle{ACM-Reference-Format}
\bibliography{references}

\end{document}